\documentclass{emulateapj}
\usepackage{lscape}

\newcommand{\lbol}{\mbox{$L_{bol}$}} % bolometric luminosity
\newcommand{\lint}{\mbox{$L_{int}$}} % internal luminosity
\newcommand{\degree}{\mbox{$^{\circ}$}}
\newcommand{\am}{\mbox{\arcmin}}
\newcommand{\as}{\mbox{\arcsec}}
\newcommand{\kms}{\mbox{km s$^{-1}$}}% km/s
\newcommand{\mjybeam}{\mbox{mJy beam$^{-1}$}}% Jy/beam

\newcommand{\um}{$\mu$m}
\newcommand{\lsun}{\mbox{L$_\odot$}}% Lsun
\newcommand{\msun}{\mbox{M$_\odot$}}% Msun
\newcommand{\water}{H$_2$O}
\newcommand{\ammonia}{\mbox{{\rm NH}$_3$}}
\newcommand{\co}{$^{12}$CO}
\newcommand{\coo}{$^{13}$CO}
\newcommand{\cooo}{C$^{18}$O}

\newcommand{\ntdp}{N$_2$D$^+$}
\newcommand{\cojone}{$^{12}$CO J $=1-0$}
\newcommand{\cojtwo}{$^{12}$CO J $=2-1$}
\newcommand{\cojthree}{$^{12}$CO J $=3-2$}
\newcommand{\coojtwo}{$^{13}$CO J $=2-1$}
\newcommand{\cooojone}{C$^{18}$O J $=1-0$}
\newcommand{\cooojtwo}{C$^{18}$O J $=2-1$}
\newcommand{\nthpjone}{N$_2$H$^+$ J $=1-0$}
\newcommand{\ntdpjthree}{N$_2$D$^+$ J $=3-2$}

\slugcomment{\footnotesize Accepted for Publication in ApJ}

\begin{document}
%%%%%%%%%%%%%%%%%% title %%%%%%%%%%%%%%%%%%%%%%%%%%%%%%%%%%%%%%%%
\title {\bf Detection of a Bipolar Molecular Outflow Driven by a Candidate First Hydrostatic Core}
\author{
Michael M.~Dunham\altaffilmark{1,2}, 
Xuepeng Chen\altaffilmark{1}, 
H\'ector G.~Arce\altaffilmark{1}, 
Tyler L.~Bourke\altaffilmark{3}, 
Scott Schnee\altaffilmark{4}, 
\& Melissa L.~Enoch\altaffilmark{5}
}

\altaffiltext{1}{Department of Astronomy, Yale University, P.O. Box 208101, New Haven, CT 06520, USA}

\altaffiltext{2}{michael.dunham@yale.edu}

\altaffiltext{3}{Harvard-Smithsonian Center for Astrophysics, 60 Garden Street, Cambridge, MA 02138, USA}

\altaffiltext{4}{National Radio Astronomy Observatory, 520 Edgemont Road, Charlottesville, VA 22903, USA}

\altaffiltext{5}{Department of Astronomy, University of California at Berkeley, 601 Campbell Hall, Berkeley, CA 94720, USA}

\begin{abstract}
We present new 230 GHz Submillimeter Array observations of the candidate first hydrostatic core Per-Bolo 58.  We report the detection of a 1.3 mm continuum source and a bipolar molecular outflow, both centered on the position of the candidate first hydrostatic core.  The continuum detection has a total flux density of 26.6 $\pm$ 4.0 mJy, from which we calculate a total (gas and dust) mass of $0.11 \pm 0.05$ \msun\ and a mean number density of $2.0 \pm 1.6 \times 10^7$ cm$^{-3}$.  There is some evidence for the existence of an unresolved component in the continuum detection, but longer-baseline observations are required in order to confirm the presence of this component and determine whether its origin lies in a circumstellar disk or in the dense inner envelope.  The bipolar molecular outflow is observed along a nearly due east-west axis.  The outflow is slow (characteristic velocity of 2.9 \kms), shows a jet-like morphology (opening semi-angles $\sim$8\degree\ for both lobes), and extends to the edges of the primary beam.  We calculate the kinematic and dynamic properties of the outflow in the standard manner and compare them to several other protostars and candidate first hydrostatic cores with similarly low luminosities.  We discuss the evidence both in support of and against the possibility that Per-Bolo 58 is a first hydrostatic core, and we outline future work needed to further evaluate the evolutionary status of this object.
\end{abstract}

%\keywords{stars: formation - stars: low-mass, brown dwarfs -  ISM: clouds - submillimeter}
%\keywords{stars: formation - ISM: clouds - submillimeter}
\keywords{stars: formation - stars: low-mass - ISM: individual objects (Per-Bolo 58) - ISM: jets and outflows}

%%%%%%%%%%%%%%%%%%%%%%%%%%%%%%%%%%%%%%%%%%%%%%%%%%%%%%%%%%%%%

\section{Introduction}\label{sec_intro}

Low-mass stars form from the gravitational collapse of dense molecular cloud cores (e.g., Shu, Adams, \& Lizano 1987; McKee \& Ostriker 2007).  Although many open questions remain, the evolutionary stages in this process are generally well understood.  Nearly all of the main stages have been observed in large samples and compared to theoretical predictions, including starless (prestellar) cores, embedded protostars, T Tauri stars, and pre-main sequence stars.  However, one stage has still not yet been definitively observed: the first hydrostatic core (hereafter FHSC).  First predicted by Larson (1969), it exists between the starless and protostellar phases.  As the molecular cloud core collapses, a central hydrostatic object (the FHSC) forms once the central density increases to the point where the central region becomes opaque to radiation ($\rho_c \ga 10^{-13}$ g cm$^{-3}$; Larson 1969), rendering the collapse adiabatic rather than isothermal.  This object continues to accrete from the surrounding core and both its mass and central temperature increase with time.  Once the temperature reaches $\sim$ 2000 K the energy liberated by accreting material dissociates H$_2$, preventing the temperature from continuing to rise to sufficiently balance gravity.  At this point the second collapse is initiated, eventually leading to the formation of the second hydrostatic core, more commonly referred to as the protostar.

Numerous theoretical studies have investigated the physical properties of first hydrostatic cores.  The maximum mass before the onset of the second collapse is generally $\sim 0.04-0.05$ \msun\ (Boss \& Yorke 1995; Masunaga, Miyama, \& Inutsuka 1998; Omukai 2007; Tomida et al.~2010), although a larger range ($0.01-0.1$ \msun) is found when the effects of rotation are included (Saigo \& Tomisaka 2006; Saigo, Tomisaka, \& Matsumoto 2008).  The lifetime of the FHSC is estimated to be short compared to the embedded phase duration of 0.54 Myr (Evans et al.~2009), but specific studies find a large range of possible lifetimes between $\sim 5 \times 10^2 - 5 \times 10^4$ yr (Boss \& Yorke 1995; Omukai 2007; Saigo, Tomisaka, \& Matsumoto 2008; Tomida et al.~2010).  Most of the variation is due to different models and assumptions concerning the mass accretion rate from the surrounding core.  The predicted internal luminosity\footnote{The internal luminosity is the luminosity of the central source and excludes luminosity arising from external heating.} of the FHSC is also very low but covers a large range ($10^{-4} - 10^{-1}$ \lsun; Masunaga, Miyama, \& Inutsuka 1998; Omukai 2007), with most of the variation again due to different mass accretion rates from the surrounding core leading to different accretion luminosities.  The radius is typically $\sim$ 5 AU (Masaunaga, Miyama, \& Inutsuka 1998), but can be much larger ($\ga 10-20$ AU) and exhibit a flattened, disk-like morphology when the effects of rotation are included (Saigo \& Tomisaka 2006; Saigo, Tomisaka, \& Matsumoto 2008).

The observational characteristics of FHSCs have also been predicted by several authors.  The emergent spectral energy distributions (SEDs) of FHSCs embedded in their parent cores have been studied by Boss \& Yorke (1995), Masunaga, Miyama, \& Inutsuka (1998), Omukai (2007), and Saigo \& Tomisaka (2011).  While the details vary from one study to the next, they all find that the radiation emitted by the FHSC is essentially completely reprocessed by the dust in the surrounding core to $50-200$ \um, with the SED characterized by emission from $10-30$ K dust and no observable emission below $30-50$ \um.  Omukai (2007) also showed that the \water\ lines emitted in the accretion shock fall below the sensitivities of current space missions but could be detected in future planned missions.  Machida, Inutsuka, \& Matsumoto (2008) showed that both the first and second cores drive outflows.  The outflow driven by the first core features slow velocities ($\la$ 5 \kms) and wide opening-angles; in contrast, the outflow driven by the second core is faster ($\sim$ 30 \kms) and well-collimated.

On the observational front, several candidate FHSCs have been found, although none have been definitively identified as true first cores.  Belloche et al.~(2006) found evidence for significant chemical evolution in the dense core Chamaeleon-MMS1, identified an associated faint infrared source in \emph{Spitzer Space Telescope} (Werner et al.~2004) observations, and noted this core harbors either a FHSC or very young protostar.  Chen et al.~(2010) detected an outflow driven by the previously unknown core L1448 IRS2E in Perseus and showed there is no associated \emph{Spitzer} infrared source down to a luminosity limit of $\sim$ 0.1 \lsun.  They suggested this is a strong FHSC candidate, although the outflow is much faster ($\sim$ 25 \kms) than expected ($\la$ 5 \kms; Machida, Inutsuka, \& Matsumoto 2008).  X.~Chen et al.~(2011, in preparation) and Pineda et al.~(2011) have detected outflows from two additional cores lacking \emph{Spitzer} infrared detections (CB 17 MMS and L1451-mm, respectively) and have suggested both as first core candidates.  Enoch et al.~(2010) detected the 70 \um\ signature expected for a FHSC in deep \emph{Spitzer} observations of the dense core Per-Bolo 58 in Perseus.  However, they also found a corresponding faint \emph{Spitzer} 24 \um\ detection in existing \emph{Spitzer} c2d legacy data (Evans et al.~2003; 2009) and argued this source could either be a FHSC or very low luminosity Class 0 protostar.

Determining the true evolutionary status of these objects is of central importance in studies of the earliest phases of low-mass star formation.  Identifying one or more as confirmed FHSCs would verify a theoretical prediction first made over 40 years ago (Larson 1969), and would open the physical properties and characteristics of FHSCs to observational study.  Furthermore, identifying one or more as confirmed very low luminosity protostars would have important implications for our current understanding of protostellar luminosities and mass accretion.  The \emph{Spiter} c2d legacy program has detected a number of very low luminosity objects (VeLLOs), protostars embedded in dense cores with internal luminosities less than or equal to 0.1 \lsun\ but generally above $\sim$ 0.05 \lsun\ (Di Francesco et al.~2007; Dunham et al.~2008), many in cores previously believed to be starless.  Recent results from this program have also shown that the protostellar luminosity distribution is strongly peaked at low luminosities (Enoch et al.~2009a; Evans et al.~2009), confirming and aggravating the classic ``luminosity problem,'' whereby protostars are underluminous compared to the accretion luminosity expected both from theoretical collapse calculations and arguments based on the minimum accretion rate necessary to form a star within the embedded phase duration (Kenyon et al.~1990; 1994; Kenyon \& Hartmann 1995).  Confirmed detections of protostars with luminosities at or below 0.01 \lsun\ would only further exacerbate the luminosity problem.

In this paper we present new 230 GHz Submillimeter Array (SMA; Ho et al.~2004) observations of Per-Bolo 58 and report the discovery of an outflow driven by this source.  We provide a brief introduction to Per-Bolo 58 in \S \ref{sec_perbolo58}, and a description of our observations in \S \ref{sec_obs}.  We present our basic results in \S \ref{sec_results}, including the detection of the continuum in \S \ref{sec_results_continuum} and the detection of \cojtwo\ and its isotopologues in \S \ref{sec_results_co}.  We analyze the properties of the molecular outflow in \S \ref{sec_outflow}, including its morphology (\S \ref{sec_outflow_morphology}) and kinematics (\S \ref{sec_outflow_kinematics}).  Finally, we discuss the implications of our results in \S \ref{sec_discussion} and summarize our findings in \S \ref{sec_summary}.

\section{Perseus Bolo 58}\label{sec_perbolo58}

Perseus Bolo 58 (R.A.~= 03 29 25.7, Decl.~= $+$31 28 16.3; hereafter Per-Bolo 58) is a dense molecular cloud core located on the northern edge of the young (proto)stellar cluster NGC 1333 in the Perseus Molecular Cloud, at an assumed distance of 250 $\pm$ 50 pc (Enoch et al.~2006), consistent with the VLBI maser parallax distance of 235 $\pm$ 18 pc for NGC 1333 recently determined by Hirota et al.~(2008).  Per-Bolo 58 was identified independently by Hatchell et al.~(2005) and Enoch et al.~(2006) in (sub)millimeter dust continuum emission surveys of Perseus using James Clerk Maxwell Telescope (JCMT) Submillimeter Common User Bolometer Array (SCUBA) 450 and 850 \um\ observations and Caltech Submillimeter Observatory (CSO) Bolocam 1.1 mm observations, respectively.  The total core mass is determined to be $0.8-1.2$ \msun\ by Enoch et al.~(2006; 2010) and 2.6 \msun\ by Hatchell et al.~(2007a).  Schnee et al.~(2010) reported a detection of Per-Bolo 58 in Combined Array for Research in Millimeter Astronomy (CARMA) 3 mm continuum observations, and from the detection calculated a total core mass of 2.4 \msun.  The differences in these mass estimates are within the uncertainties introduced by different gas-to-dust ratio, dust opacity, and dust temperature assumptions.

\begin{deluxetable*}{lcccccc}
\tabletypesize{\scriptsize}
\tablewidth{0pt}
\tablecaption{\label{tab_smaobs}SMA Observations of Perseus Bolo 58}  
\tablehead{
\colhead{} & \colhead{$\nu$} & \colhead{Beam FWHM} & \colhead{Beam PA\tablenotemark{a}} & \colhead{Bandwidth} & \colhead{$\delta$V\tablenotemark{b}} & \colhead{$1\sigma$ rms\tablenotemark{c}}\\
\colhead{Line} & \colhead{(GHz)} & \colhead{(Arcseconds)} & \colhead{(Degrees)} & \colhead{(GHz)} & \colhead{(\kms)} & \colhead{(\mjybeam)}}
\startdata
\cojtwo\ & 230.53797 & $2.92 \times 2.49$ & $-$86.0 & 0.082 & 0.53 & 42 \\
\coojtwo\ & 220.39868 & $3.15 \times 2.63$ & $-$74.5 & 0.082 & 0.53 & 33 \\
\cooojtwo\ & 219.56037 & $3.16 \times 2.64$ & $-$74.6 & 0.082 & 0.26 & 46 \\
\hline
Continuum & 225.67836 & $3.01 \times 2.54$ & $-$80.0 & 5.330 & \nodata & 0.39 \\
\enddata\\
\tablenotetext{a}{Position angle of the long axis of the beam, measured east (counterclockwise) from north.}
\tablenotetext{b}{Width of each channel in \kms.}
\tablenotetext{c}{For lines, the mean of the 1$\sigma$ rms of the spectrum at each spatial position with the spectral resolution given in the previous column.  For the continuum, the $1\sigma$ rms of the continuum intensity.}
\end{deluxetable*}

Per-Bolo 58 was identified as starless by both Enoch et al.~(2008) and Hatchell et al.~(2007a) based on the lack of an associated \emph{Spitzer} infrared source.  It was also identified as starless by Hatchell et al.~(2007b) and Hatchell \& Dunham (2009) based on the nondetection of an outflow in single-dish JCMT \cojthree\ data.  Curtis et al.~(2010) also reported the nondetection of an outflow in their independent JCMT \cojthree\ data.  Arce et al.~(2010) detected a redshifted COMPLETE\footnote{COordinated Molecular Probe Line Extinction Thermal Emission; Ridge et al.~(2006)} Perseus Outflow Candidate (CPOC 19) in their single-dish Five College Radio Astronomy Observatory (FCRAO) \cojone\ data beginning $\sim$1\am\ northeast of Per-Bolo 58 and extending several arcminutes farther to the northeast.  However, they identified IRAS 03262$+$3123 as the most likely driving source.  Furthermore, as we will show below in \S \ref{sec_outflow_morphology}, the position and morphology of CPOC 19 indicate it is unlikely to be associated with Per-Bolo 58.

Enoch et al.~(2010) obtained deep 70 \um\ observations of this object as part of a program attempting to detect the FHSC by targeting cores classified as starless but meeting various criteria suggesting they may already be collapsing.  Per-Bolo 58 was detected with a low 70 \um\ flux density (65 mJy) consistent with theoretical predictions for FHSCs.  They also noted an associated 4$\sigma$ 24 \um\ detection in existing, shallower \emph{Spitzer} c2d data first published by J\o rgensen et al.~(2006)\footnote{This 4$\sigma$ 24 \um\ detection was not reported in the list of candidate young stellar objects (YSOs) reported by J\o rgensen et al.~(2006), but was present in the full source catalog assembled by the c2d team (Evans et al.~2007).} above the flux level predicted for a FHSC surrounded by a spherically symmetric envelope.  From continuum radiative transfer models, Enoch et al.~determined this source has an internal luminosity of \lint\ $\sim$ 0.012 \lsun.  They showed that the observed SED could be fit by either a very low luminosity protostar or a first hydrostatic core with a spherical and/or outflow cavity to reduce the 24 \um\ opacity, and were unable to rule out the possibility that Per-Bolo 58 is a low luminosity protostar based on the SED alone.

%AKA HRF64.

\section{Observations}\label{sec_obs}

One track of observations of Per-Bolo 58 were obtained with the SMA on 2010 November 16 in the compact configuration with 8 antennas, providing projected baselines ranging from $5-76$ m.  The observations were obtained with the 230 GHz receiver and included 4 GHz of bandwidth per sideband, with 12 GHz spacing between the centers of the two sidebands.  The correlator was configured such that the lower sideband (LSB) covered approximately $216.8-220.8$ GHz while the upper sideband (USB) covered aproximately $228.8-232.8$ GHz, providing simultaneous observations of the \co, \coo, \cooojtwo, \ntdpjthree, and SiO J $=5-4$ lines.  The lines were observed with either 256 (\coo, SiO) or 512 (\co, \cooo, \ntdp) channels in the 104 MHz bands, providing channel separations of 0.53 and 0.26 \kms, respectively.  The remaining bands were used to measure the 1.3 mm continuum with a total bandwidth of 5.33 GHz.

The weather conditions were extremely good for 230 GHz observations.  Throughout the observations the zenith opacity at 225 GHz varied between $\sim 0.05-0.1$, and the system temperature was typically $\sim 120$K, ranging from $80-240$ K depending on elevation.  Regular observations of the quasar 3C84 were interspersed with those of Per-Bolo 58 for gain calibration.  Uranus and 3C84 were used for passband calibration, and Uranus was used for absolute flux calibration.  We estimate a 15\% uncertainty in the absolute flux calibration by comparing the measured fluxes of the calibrators from our calibrated data with those in the SMA calibrator database\footnote{Available at http://sma1.sma.hawaii.edu/callist/callist.html} for the same observation date.  The data were inspected, flagged, and calibrated using the MIR software package\footnote{Available at https://www.cfa.harvard.edu/$\sim$cqi/mircook.html} and imaged, cleaned, and restored using the Multichannel Image Reconstruction, Image Analysis, and Display (MIRIAD) software package configured for the SMA\footnote{Available at http://www.cfa.harvard.edu/sma/miriad/}.  Imaging was performed with a robust uv weighting parameter of $+$1, found to provide the best compromise between resolution and sensitivity for these data.  The final maps were re-gridded onto 0.5\as\ pixels.

Only the continuum and CO (and its isotopologues) are detected and discussed in this paper.  Table \ref{tab_smaobs} lists, for each of these observations, the frequency of observation, the synthesized beam size and orientation, the total bandwidth, the channel separation (for the lines), and the measured $1\sigma$ rms.  For the continuum, the 1$\sigma$ rms is determined by calculating the standard deviation of all off-source pixels.  For the spectra, the 1$\sigma$ rms is determined by calculating, for each pixel, the standard deviation of the intensity in each spectral channel outside of the velocity range $0-20$ \kms, and then calculating the mean standard deviation over all pixels.

\section{Results}\label{sec_results}

\subsection{Continuum}\label{sec_results_continuum}

\begin{figure}
\epsscale{1.0}
\plotone{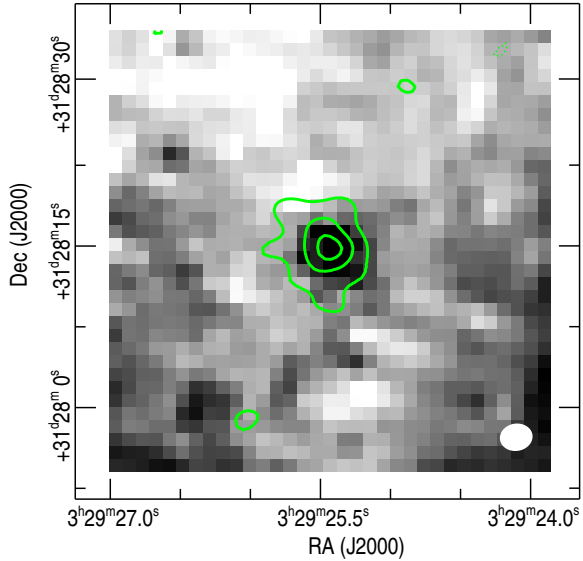}
\vspace{0.2 in}
\caption{\label{fig_mips_cont}\emph{Spitzer} c2d 24 \um\ image of Per-Bolo 58.  The grayscale is inverted and displayed in a linear stretch with the minimum and maximum intensities set to 36.5 (white) and 37.1 (black) MJy sr$^{-1}$, respectively.  The green contours show the SMA 1.3 mm continuum intensity.  The solid contours show positive intensity contours starting at 3$\sigma$ and increasing by 7$\sigma$ and the dotted contours show negative intensity contours starting at $-$3$\sigma$ and decreasing by 7$\sigma$, where the 1$\sigma$ rms in the continuum image is 0.39 mJy beam$^{-1}$.  The two $3\sigma$ peaks southeast and northwest of Per-Bolo 58 are sidelobes incompletely removed by the cleaning process.  Only one $-3\sigma$ peak is present in the northwest corner of the image.  The synthesized beam size and shape of the SMA continuum observations is shown by the white filled oval in the lower right.}
\end{figure}

Figure \ref{fig_mips_cont} displays, in inverted grayscale, the \emph{Spitzer} c2d 24 \um\ image of Per-Bolo 58, showing the weak detection at this wavelength.  Overlaid are contours showing the 1.3 mm SMA continuum intensity.  Per-Bolo 58 is clearly detected in the continuum.  Using the MIRIAD task \emph{imfit} we fit an elliptical Gaussian to the detection and report a peak position of R.A.~= 03 29 25.46, Decl.~= $+$31 28 15.0, as listed in Table \ref{tab_continuum}.  The uncertainties in this position from the fit are less than 0.2\as\ in both Right Ascension and Declination.  As evident from Figure \ref{fig_mips_cont}, this position shows excellent agreement with the position of the infrared source detected by Enoch et al.~(2010).  It also agrees to within $1.5\as$ with the position reported by Schnee et al.~(2010) from their 3 mm CARMA detection.  Our measured peak position is offset by 3.3\as\ to the southwest of the Bolocam 1.1 mm single-dish continuum position reported by Enoch et al.~(2010).  As the Bolocam data includes an absolute pointing uncertainty of 7\as, we do not consider the offset to be significant.

Other properties derived from the elliptical Gaussian fit to the detected source, including the peak flux density, the total flux density, the deconvolved source size, and the deconvolved source position angle, are also reported in Table \ref{tab_continuum}.  The total flux density and effective radius\footnote{The effective radius is defined as the geometric mean of the semimajor and semiminor axes} of the continuum source are 26.6 $\pm$ 4.0 mJy and 2.3 $\pm$ 0.2\as\ (575 $\pm$ 125 AU at the assumed distance of 250 $\pm$ 50 pc), respectively.  For comparison, Enoch et al.~(2006) measured a total flux density and effective radius of $330 \pm 30$ mJy and $21 \pm 1$\as\ ($5250 \pm 1080$ AU at the assumed distance of $250 \pm 50$ pc), respectively, in their single-dish Bolocam 1.1 mm continuum data.  Scaling this 1.1 mm flux density to 1.3 mm assuming optically thin emission with an opacity power-law index of $\beta = 1.5$ gives an expected 1.3 mm flux density of $171 \pm 16$ mJy\footnote{The uncertainty in the expected 1.3 mm flux density includes only the uncertainty in the Bolocam flux density and does not include any component from the uncertain dust opacity power-law assumption.  Power-law indices of 1.0 and 2.0 would give expected 1.3 mm flux densities of $187 \pm 17$ and $155 \pm 14$, respectively.}.  This factor of 6 discrepancy in flux density and factor of 9 difference in effective radius between the SMA and Bolocam observations is easily explained by the fact that less than 3\% of the total uv pointings from our compact configuration SMA observations are located at projected baselines $<$ 10 k$\lambda$ (corresponding to angular scales $>$ 21\as), thus these data filter out the majority of the emission from the large-scale extended core and are instead sensitive only to the emission from the inner, compact envelope and/or circumstellar disk.

%\begin{landscape}
\begin{deluxetable*}{cccccccc}
\tabletypesize{\scriptsize}
\tablewidth{0pt}
\tablecaption{\label{tab_continuum}Elliptical Gaussian Fit to the 1.3 mm Continuum Detection}  
\tablehead{
\colhead{Peak R.A.} & \colhead{Peak Decl.} & \colhead{Peak Flux Density}        & \colhead{Total Flux Density} & \colhead{Source Size\tablenotemark{a}} & \colhead{Source P.A.\tablenotemark{a}} & \colhead{$M$}     & \colhead{$n$} \\
\colhead{(J2000)}   & \colhead{(J2000)}    & \colhead{(mJy beam$^{-1}$)} & \colhead{(mJy)}      & \colhead{(\as)}       & \colhead{(degrees)}   & \colhead{(\msun)} & \colhead{(cm$^{-3}$)}}
\startdata
03:29:25.46 & $+$31:28:15.0 & 6.9 $\pm$ 1.1 & 26.6 $\pm$ 4.0        & 5.3 $\pm$ 0.4 $\times$ 4.0 $\pm$ 0.3 & 19.0 $\pm$ 16.2  & 0.11 $\pm$ 0.05 & $2.0 \pm 1.6 \times 10^7$
\enddata\\
\tablenotetext{a}{Deconvolved with the beam.}
\end{deluxetable*}
%\end{landscape}

\begin{figure}
\epsscale{1.0}
\plotone{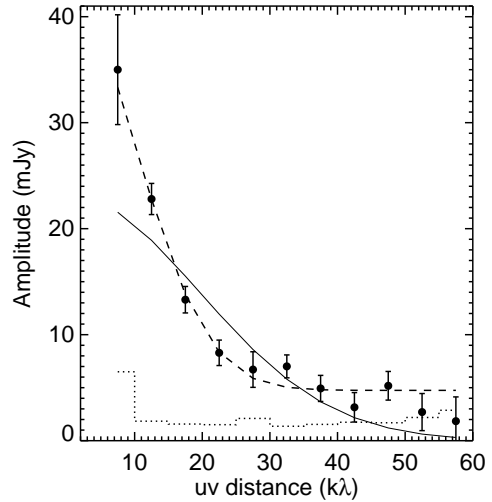}
\caption{\label{fig_visibilities}SMA 1.3 mm continuum visibility amplitude versus \emph{uv}-distance for Per-Bolo 58.  The amplitudes are calculated with the MIRIAD task \emph{uvamp}, which bins the data in annuli according to uv distance and calculates the vector averaged amplitude in each annulus.  The annuli are spaced by 5 k$\lambda$, and the error bars give the standard deviation in each annulus.  The dotted line shows the expectation value for zero signal.  The solid line shows the best-fit Gaussian function, while the dashed line shows the best-fit Gaussian$+$constant function.  In both cases the Gaussian fits were calculated using the IDL procedure \emph{gaussfit} (see text for more details).}
\end{figure}

Figure \ref{fig_visibilities} shows the SMA 1.3 mm continuum visibility amplitudes for Per-Bolo 58.  There may be evidence of a compact, unresolved component seen in baselines longer than $\sim$ 30 k$\lambda$.  To investigate this possibility, we used the IDL procedure \emph{gaussfit} to fit both a Gaussian and a Gaussian plus constant offset to the data assuming Gaussian errors.  As shown in Figure \ref{fig_visibilities}, these fits indicate that a Gaussian plus constant offset (reduced $\chi^2=0.9$, with a best-fit offset of 4.7 mJy) gives a significantly better fit than a Gaussian alone (reduced $\chi^2=4.3$).  We note here that if we implement \emph{gaussfit} with Poisson ($\sqrt{N}$) errors instead of Gaussian errors, appropriate for the low signal-to-noise regime, a Gaussian plus constant offset (reduced $\chi^2=0.5$, with a best-fit offset of 3.3 mJy) still gives a better fit by eye than a Gaussian alone (reduced $\chi^2=1.5$), although the quantitative difference between the quality of the two fits is small.

Based on these results we conclude that the presence of an unresolved component is possible but only marginally supported by the data, especially considering that Gaussians are likely poor representations of the true emission profiles.  If real, this unresolved component could represent a circumstellar disk or the flattened, disk-like structure of a rapidly rotating FHSC (e.g., Saigo \& Tomisaka 2006; Saigo, Tomisaka, \& Matsumoto 2008).  However, at the distance to Perseus, projected baselines ranging from $30-50$ k$\lambda$ correspond to spatial scales ranging from $1000-1700$ AU, thus the unresolved component could also simply correspond to the dense inner envelope.  Future observations in more extended configurations providing longer baselines must be combined with radiative transfer models of a star+disk+envelope system fit to both the visibility amplitudes and the observed SED (e.g., J\o rgensen et al.~2005; Enoch et al.~2009b) in order to fully separate inner envelope and disk components and confirm the presence of this unresolved component.

The last two columns of Table \ref{tab_continuum} list the mass and density derived from the SMA continuum detection.  The mass is calculated as

\begin{equation}\label{eq_dustmass}
M = 100 \frac{d^2 S_{\nu}}{B_{\nu}(T_D) \kappa_{\nu}} \quad ,
\end{equation}
where $S_{\nu}$ is the total flux density, $B_{\nu}(T_D)$ is the Planck function at the isothermal dust temperature $T_D$, $\kappa_{\nu}$ is the dust opacity, $d = 250 \pm 50$ pc, and the factor of 100 is the assumed gas-to-dust ratio.  We adopt the dust opacities of Ossenkopf \& Henning (1994) appropriate for thin ice mantles after $10^5$ yr of coagulation at a gas density of $10^6$ cm$^{-3}$ (OH5 dust), giving $\kappa_{\nu} = 0.864$ cm$^2$ gm$^{-1}$ at the frequency of the continuum observations.  For comparison, Enoch et al.~(2006; 2010) and Schnee et al.~(2010) also adopt OH5 dust opacities at the frequencies of their observations to calculate masses, while Hatchell et al.~(2007a) adopt a dust opacity 1.5 times higher than that of OH5 dust at the frequency of their observations.  With the total flux density listed in Table \ref{tab_continuum} and an assumed $T_D = 10$ K, Equation \ref{eq_dustmass} gives a total mass of 0.11 $\pm$ 0.05 \msun.  Note that the uncertainty of 0.05 \msun\ only includes the statistical uncertainty in the flux density and distance.  The true uncertainty is likely dominated by the dust temperature and opacity assumptions.  For example, the dust opacity at 1.3 mm can vary by factors of $\sim 2-4$ depending on which dust opacity model is adopted (e.g., Shirley et al.~2005; 2011), directly leading to factors of $\sim 2-4$ variation in the mass.  Furthermore, the exact dust temperature is uncertain.  Rosolowsky et al.~(2008) identified two velocity components in their \ammonia\ observations (see further discussion of the two components in \S \ref{sec_results_co} below) with derived kinetic temperatures of 10.3 and 10.5 K, respectively, in good agreement with our assumption of 10 K.  An assumed temperature of 20 K would change the derived mass to 0.04 $\pm$ 0.02 \msun.

The mean number density, $n$, is calculated assuming spherical symmetry as

\begin{equation}\label{eq_numberdens}
n = \frac{3}{4 \pi \mu m_H} \frac{M}{r^3_{eff}}
\end{equation}
where $M$ is the mass, $r_{eff}$ is the effective radius, $m_{H}$ is the hydrogen mass, and $\mu$ is the mean molecular weight per free particle.  With $M$ and $r_{eff}$ as given above and $\mu = 2.37$ for gas that is 71\% by mass hydrogen, 27\% helium, and 2\% metals (Kauffmann et al.~2008), we calculate $n= 2.0 \pm 1.6 \times 10^7$ cm$^{-3}$, where the uncertainty includes the statistical uncertainties in the mass and effective radius as calculated above.

\subsection{CO and Isotopologues}\label{sec_results_co}

\begin{figure}
\epsscale{1.0}
\plotone{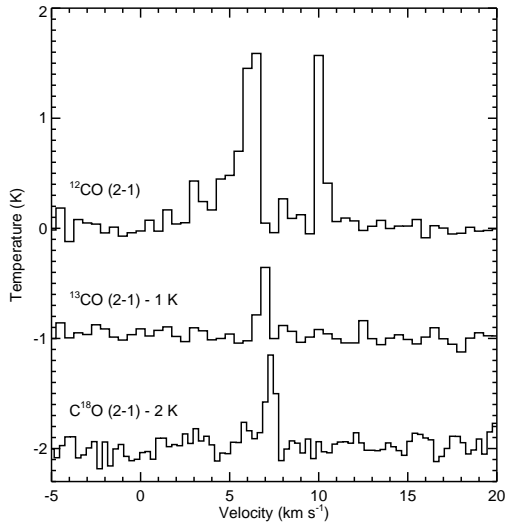}
\caption{\label{fig_spectra}\co, \coo, and \cooojtwo\ spectra of Per-Bolo 58 averaged over the central 4\as\ relative to the position of the continuum source.  The \coo\ and \cooo\ spectra are shifted down by 1 and 2 K, respectively, for display purposes.}
\end{figure}

Figure \ref{fig_spectra} shows the \co, \coo, and \cooojtwo\ spectra of Per-Bolo 58 averaged over the central 4\as\ relative to the position of the continuum source.  The \cooo\ line is clearly detected as a narrow line centered at $\sim$ 7.3 \kms.  Gaussian fits to the individual 0.5\as\ pixels within this central region using the IDL procedure \emph{gaussfit} yield central velocities between 7.2 and 7.5 \kms, with both a mean and median of 7.3 \kms.  This dispersion in velocity of 0.3 \kms\ is approximately within one spectral resolution element (0.26 \kms; Table \ref{tab_smaobs}), thus we adopt 7.3 \kms\ as the core rest velocity of Per-Bolo 58.  For comparison, Hatchell \& Dunham (2009) derived a core velocity of 7.9 \kms\ based on single-dish FCRAO \cooojone\ data with a spatial and spectral resolution of 46\as\ and 0.25 \kms, respectively, and Kirk, Johnstone, \& Tafalla (2007) found a velocity of 7.53 (7.54) \kms\ based on single-dish IRAM 30 m \cooojtwo\ (\nthpjone) data with a spatial and spectral resolution of 11\as\ (25\as) and 0.05 \kms\ (0.05 \kms), respectively.  The range of values between $7.3-7.9$ \kms\ may be explained by the fact that Rosolowsky et al.~(2009) argued that two velocity components, one at 7.4 and one at 7.77 \kms, are necessary to fit their \ammonia\ observations.  Further detections of dense gas tracers, particularly at the high spatial resolution provided by interferometers such as the SMA, are needed to further investigate the true core velocity.

The Gaussian fits to the individual pixels within the central 4\as\ give full-width half-maximum (FWHM) linewidths varying between 0.3 and 1.4 \kms, with a mean value of 0.7 \kms.  The lines are generally spread over only $2-4$ velocity channels and are thus only marginally resolved.  Purely thermal gas at 10 K would produce FWHM linewidths of 0.12 \kms\ and lines confined to a single channel given the channel width of $\sim 0.3$ \kms.  While nonthermal components thus appear to be present, higher density tracers are required to rule out contamination by emission from outflowing gas (see below).  No significant ($> 5\sigma$) \cooo\ emission is detected outside of this central region.

The \coo\ line is also clearly detected, centered at $\sim$ 7 \kms.  Gaussian fits to the individual pixels within this central region, again using the IDL procedure \emph{gaussfit}, yield central velocities between 6.8 and 7.3 \kms, with both a mean and median of 6.9 \kms.  We consider the \cooo\ line a better measure of the true core rest velocity than the \coo\ line since the latter is more likely to include emission from outflowing gas and also more likely to be spatially extended and thus partially filtered out.  We detect \coo\ emission near the core velocity at levels greater than $5\sigma$ over the full extent of our map, but the emission is sporadic and not clearly tracing either the central core or outflow axis (see below), and no significant emission is detected at $|v-v_{\rm core}| > 2$ \kms.  This emission likely represents the most compact components of extended \coo\ cloud emission, most of which is resolved out.  Shorter baseline data are necessary to fully reconstruct a map.

\begin{figure}
\epsscale{1.0}
\plotone{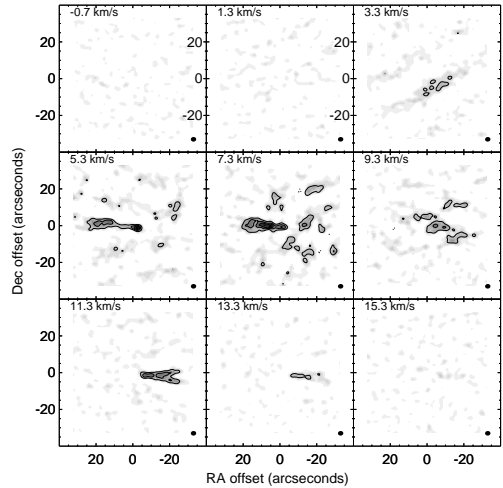}
\caption{\label{fig_12co21_cmap}SMA \cojtwo\ channel maps of Per-Bolo 58 with a velocity resolution of 2 \kms, with the velocities marked in the top left of each panel.  The gray scale shows intensity in units of Jy beam$^{-1}$ with a linear scaling ranging from 0.0 Jy beam$^{-1}$ (white) to 0.50 Jy beam$^{-1}$ (black).  Overplotted with solid lines are positive intensity contours starting at 5$\sigma$ and increasing by 5$\sigma$ and with dotted lines are negative intensity contours starting at $-$5$\sigma$ and decreasing by 5$\sigma$, where the 1$\sigma$ rms in the 2 \kms\ channels is 0.022 Jy beam$^{-1}$.  The synthesized beam size and shape of the \cojtwo\ observations is shown by the black filled oval in the lower right of each panel.  The central (0,0) position is that reported by Enoch et al.~(2010) (the phase center of the observations).}
\end{figure}

Finally, emission from \co\ is also clearly detected.  The spectra show two peaks blueshifted and redshifted relative to the core velocity and no emission at the core velocity, classic signatures of an outflow observed with an interferometer (e.g., Bourke et al.~2005) with the emission from the core itself resolved out.  Figure \ref{fig_12co21_cmap} displays channel maps of the \co\ data at a velocity resolution of 2 \kms.  Blueshifted emission between $\sim 4-7$ \kms\ extends to the east (including in the central velocity channel due to the 2 \kms\ channel widths) whereas redshifted emission between $\sim 10-13$ \kms\ extends to the west.  With our assumed core velocity of 7.3 \kms, the blueshifted emission appears to be moving slower than the redshifted emission.  We will return to this point in \S \ref{sec_outflow_kinematics}.  No outflow emission is seen beyond $|v-v_{\rm core}| \sim 7$ \kms.

\section{Molecular Outflow}\label{sec_outflow}

\subsection{Morphology}\label{sec_outflow_morphology}

\begin{figure}
\epsscale{1.0}
\plotone{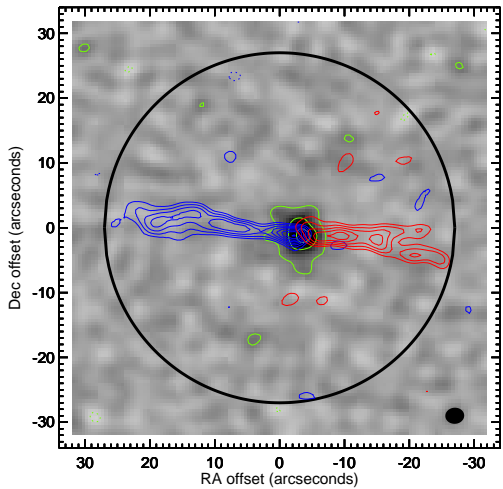}
\caption{\label{fig_12co21_wings}SMA 1.3 mm continuum image of Per-Bolo 58.  The grayscale is inverted and displayed in a linear stretch ranging from -2.73 mJy beam$^{-1}$ ($-7\sigma$, white) to 4.68 mJy beam$^{-1}$ ($12\sigma$, black).  The green contours show the continuum intensity starting at 3$\sigma$ and increasing by 7$\sigma$, where the 1$\sigma$ rms in the continuum image is 0.39 mJy beam$^{-1}$.  The blue contours show blueshifted \cojtwo\ emission integrated from 0.3 to 7.3 \kms, while the red contours show redshifted emission integrated from 7.3 to 14.3 \kms.  The solid blue and red contours start at $3\sigma$ and increase by $1\sigma$ whereas the dotted blue and red contours start at $-$3$\sigma$ and decrease by 1$\sigma$, where the $1\sigma$ rms in both the integrated red and blue maps is 0.16 Jy beam$^{-1}$ \kms.  The black circle shows the primary beam of the SMA at 230 GHz.  The synthesized beam size and shape of the \co\ observations is shown by the black filled oval in the lower right.  The central (0,0) position is that reported by Enoch et al.~(2010) (the phase center of the observations).}
\end{figure}

Figure \ref{fig_12co21_wings} shows integrated blueshifted and redshifted emission contours overlaid on the 1.3 mm continuum image.  The east-west outflow is clearly seen, with the blueshifted lobe extending almost due east, the redshifted lobe extending almost due west, and very little overlap between the lobes.  From this observed morphology, it is clear that the candidate outflow feature CPOC 19 reported by Arce et al.~(2010), with redshifted emission extending to the northeast of Per-Bolo 58, is not related to Per-Bolo 58 and is likely a high-velocity feature associated with another source in NGC 1333.  Enoch et al.~(2010) noted the presence of faint nebulosity detected in the \emph{Spitzer} Infrared Array Camera (IRAC; Fazio et al.~2004) band 2 (4.5 \um) image and suggested this emission may be tracing a weak outflow since this band overlaps with emission from molecular hydrogen heated by outflow shocks (e.g., Noriega-Crespo et al.~2004; De Buizer \& Vacca 2010).  This nebulosity extends a few arcseconds to the east of Per-Bolo 58 and is aligned with the blue outflow lobe, thus we confirm that it is indeed likely associated with the outflow.

We measure a position angle of 85\degree\ (measured east from north) for the axis connecting both lobes and passing through the position of the infrared source.  The outflow shows a narrow, jet-like morphology, with opening semi-angles of $\sim$ 8\degree\ for both the blue and red lobes.  These opening semi-angles are best treated as upper limits since the lobes are only marginally resolved in the direction perpendicular to the flow (total extents of $\sim$ 7\as, compared to the $2.9 \times 2.5$\as\ beam).  The blue lobe extends 27\as\ ($6750 \pm 1350$ AU at $250 \pm 50$ pc) from the infrared source whereas the red lobe only extends 22\as\ ($5500 \pm 1100$ AU at $250 \pm 50$ pc); we list these values in the Table \ref{tab_outflow} along with other quantities described in \S \ref{sec_outflow_kinematics} below.  However, given that the infrared source is offset by 3.3\as\ from the phase center (\S \ref{sec_results_continuum}), both lobes extend nearly to the half-power radius of the primary beam (HWHM $\sim$ 27\as\ at 230 GHz).  The lobe sizes reported here are thus lower limits to the true sizes.

Given the morphology of this outflow, in particular the two distinct, bipolar lobes showing either entirely blueshifted or redshifted emission, the geometry of the Per-Bolo 58 system is best described by case 2 of Cabrit \& Bertout (1986): the system inclination\footnote{An inclination of 0\degree\ corresponds to a face-on system, whereas an inclination of 90\degree\ corresponds to an edge-on system.} is greater than or equal to the opening semi-angle of the outflow, and the sum of the inclination and outflow semi-opening angle is less than or equal to 90\degree.  However, with such a jet-like morphology with a very small opening semi-angle of $\sim$ 8\degree, we are only able to rule out extreme edge-on and face-on inclinations.  In general we are unable to set strong constraints on the inclination, a point we will return to in \S \ref{sec_discussion} below.

\subsection{Kinematics}\label{sec_outflow_kinematics}

We calculate, separately for the blue and red lobes and for the combined outflow, the mass of the outflow, $M_{flow}$, and its kinematic and dynamic properties (kinetic energy [$E_{flow}$], momentum [$P_{flow}$], characteristic velocity [$v=P_{flow}/M_{flow}$], dynamical time [$\tau_{d}=R_{lobe}/v$], mechanical luminosity [$L_{flow}=E_{flow}/\tau_{d}$], and force [$F_{flow}=P_{flow}/\tau_{d}$]) in the standard manner (e.g., Cabrit \& Bertout 1986, 1990; see also Dunham et al.~2010).  The results are listed in Table \ref{tab_outflow} using a map corrected for primary beam attenuation.  The mass (and all other properties that depend on mass) are calculated assuming LTE and do not include corrections for inclination or optical depth.  Thus, combined with the facts that the outflow likely extends beyond the primary beam (\S \ref{sec_outflow_morphology}) and that these observations will filter out any extended component to the outflow, our reported values are lower limits only.  Because we only calculate lower limits, we assume the gas temperature that minimizes the LTE calculation of mass, 17.6 K (see Dunham et al.~2010 for details).  The actual temperatures of outflowing gas may be significantly higher ($50-200$ K; Hatchell et al.~1999; van Kempen et al.~2009a, 2009b), although this remains a significant unknown, especially for very low luminosity sources like Per-Bolo 58.  The true values of the quantities listed in Table \ref{tab_outflow} may be higher by more than an order of magnitude depending on the true values of the outflowing gas temperature and optical depth in the \cojtwo\ line and the fraction of total outflow emission missing from our map (see, e.g., Bourke et al.~2005, Chen et al.~2010, Arce et al.~2010, Curtis et al.~2010, and Dunham et al.~2010 for a few of many recent discussions on calculating outflow properties).

\begin{deluxetable*}{lllll}
\tabletypesize{\scriptsize}
\tablewidth{0pt}
\tablecaption{\label{tab_outflow}Outflow Properties}
\tablehead{\colhead{Quantity} & \colhead{Unit} & \colhead{Blue Lobe} & \colhead{Red Lobe} & \colhead{Combined}}
\startdata
Lobe Size ($R_{lobe}$)               & AU                    & $6.8 \times 10^{3}$  & $5.5 \times 10^{3}$  & $6.2 \times 10^{3}$  \\
Outflow Mass ($M_{flow}$)            & \msun                 & $1.1 \times 10^{-4}$ & $0.9 \times 10^{-4}$ & $2.0 \times 10^{-4}$ \\
Outflow Momentum ($P_{flow}$)        & \msun\ \kms           & $2.5 \times 10^{-4}$ & $3.1 \times 10^{-4}$ & $5.6 \times 10^{-4}$ \\
Outflow Kinetic Energy ($E_{flow}$)  & ergs                  & $2.3 \times 10^{40}$ & $2.8 \times 10^{40}$ & $5.1 \times 10^{40}$ \\
Outflow Luminosity ($L_{flow}$)      & \lsun                 & $1.2 \times 10^{-5}$ & $2.6 \times 10^{-5}$ & $4.2 \times 10^{-5}$ \\
Outflow Force ($F_{flow}$)           & \msun\ \kms\ yr$^{-1}$ & $1.6 \times 10^{-8}$ & $3.5 \times 10^{-8}$ & $5.6 \times 10^{-8}$  \\
Characteristic Velocity ($v_{flow}$) & \kms\                 & 2.4                  & 3.5                 & 2.9                  \\
Dynamical Time ($\tau_{d}$)         & yr                     & $1.6 \times 10^{4}$  & $0.9 \times 10^{4}$  & $1.0 \times 10^{4}$  \\
\enddata \\
\end{deluxetable*}

While the morphology of the blue and red lobes argues strongly for the presence of an outflow, with such a low characteristic velocity (2.9 \kms) it is important to rule out bound motion as being responsible for the kinematic signatures interpreted as an outflow.  For gas at 2.9 \kms\ to be bound at 6200 AU (the average observed size of the two lobes) requires a mass of at least 29 \msun, a factor of $\sim 15-30$ times higher than the total core mass.  We thus definitively eliminate bound motion as being responsible for the observed blue and red lobes.

The mass of outflowing gas in the blue and red lobes are comparable ($\sim 1 \times 10^{-4}$ \msun), but the characteristic velocity of the red lobe is faster by 1.1 \kms\ (46\%), leading to larger values of all other quantities except for the dynamical time, which is inversely proportional to velocity and thus smaller.  In particular, the force required to drive the red lobe is a factor of 2 higher than that required to drive the blue lobe.  Such a force mismatch could, in principle, accelerate the central object out of the dense core from which it is forming.  In reality, however, this acceleration is very small; a simple calculation assuming a central mass of 0.01 \msun\footnote{Approximately the amount of mass that would accrete in the dynamical time of the outflow ($\sim 10^4$ yr) at a standard rate of $\sim 10^{-6}$ \msun\ yr$^{-1}$.} shows that it would take 0.54 Myr for the central object to move 2.3\as\ (575 AU at 250 pc, the effective radius of the SMA continuum source).  This time is comparable to the total duration of the embedded phase ($0.44-0.54$ Myr; Evans et al.~2009).  Furthermore, if we shift the central velocity of the core from our assumed value of 7.3 \kms\ to 7.9 \kms, as given by Hatchell \& Dunham (2009), the forces required to drive the two lobes agree to within 16\% and the mismatch is negligible.  Detections of higher density tracers than \cooo, used by both Hatchell \& Dunham (2009) and by us to determine the core velocity, are required to better pin down the true velocity of Per-Bolo 58.

\section{Discussion}\label{sec_discussion}

Between Per-Bolo 58 (Enoch et al.~2010; this work), L1448 IRS2E (Chen et al.~2010), and L1451-mm (Pineda et al.~2011), three candidate FHSCs have been identified in Perseus.  Enoch et al.~(2009a) identified 66 embedded protostars in Perseus based on the combination of complete surveys of the cloud with the \emph{Spitzer} c2d Legacy project (J\o rgensen et al.~2006; Rebull et al.~2006) and Bolocam 1.1 mm continuum observations (Enoch et al.~2006).  Assuming the total duration of the embedded stage is 0.54 Myr (Evans et al.~2009) and the lifetime of the FHSC stage is between $5 \times 10^2 - 5 \times 10^4$ yr (Boss \& Yorke 1995; Omukai 2007; Saigo, Tomisaka, \& Matsumoto 2008; Tomida et al.~2010), there should be $<1-6$ FHSCs in Perseus.  The current number of candidates (three) is near the maximum of this range, suggesting a relatively long duration of the FHSC stage (at least $2.5 \times 10^4$ yr) if all three are in fact first cores.

With a characteristic velocity of 2.9 \kms\ and a maximum velocity of $\sim$ 7 \kms, the Per-Bolo 58 outflow is relatively slow compared to other collimated outflows from young, embedded objects (e.g., Gueth \& Guilloteau 1999; Lee et al.~2000; Arce \& Sargent 2006).  Such slow velocities agree with the prediction by Machida, Inutsuka, \& Matsumoto (2008) that the FHSC drives a slow ($\la 5$ \kms) outflow.  The outflow also features a collimated, jet-like appearance, with an opening semi-angle of $\sim$ 8\degree\ and a lobe extent-to-width ratio ($E/W$) $\geq 3-4$.  Machida, Inutsuka, \& Matsumoto (2008) predict that the outflows driven by first cores have very wide opening angles and constant $E/W \sim 2.2-2.5$, whereas those driven by second cores (protostars) are more collimated with $E/W$ starting around 5 and increasing with time.  The outflow driven by Per-Bolo 58 appears to feature a morphology between these two cases but more consistent with a protostellar rather than FHSC outflow, although, as noted in \S \ref{sec_results_continuum}, less than 3\% of the total uv pointings from our compact configuration SMA observations are located at projected baselines $<$ 10 k$\lambda$ (corresponding to angular scales $>$ 21\as).  Future observations providing shorter baselines and/or deeper single-dish data than published by Hatchell \& Dunham (2009), Curtis et al.~(2010), and Arce et al.~(2010) should be pursued to determine whether or not a wider component to the outflow more extended than $\sim 21\as$ exists and is resolved out in the current SMA data.

\begin{figure*}
\epsscale{1.0}
\plotone{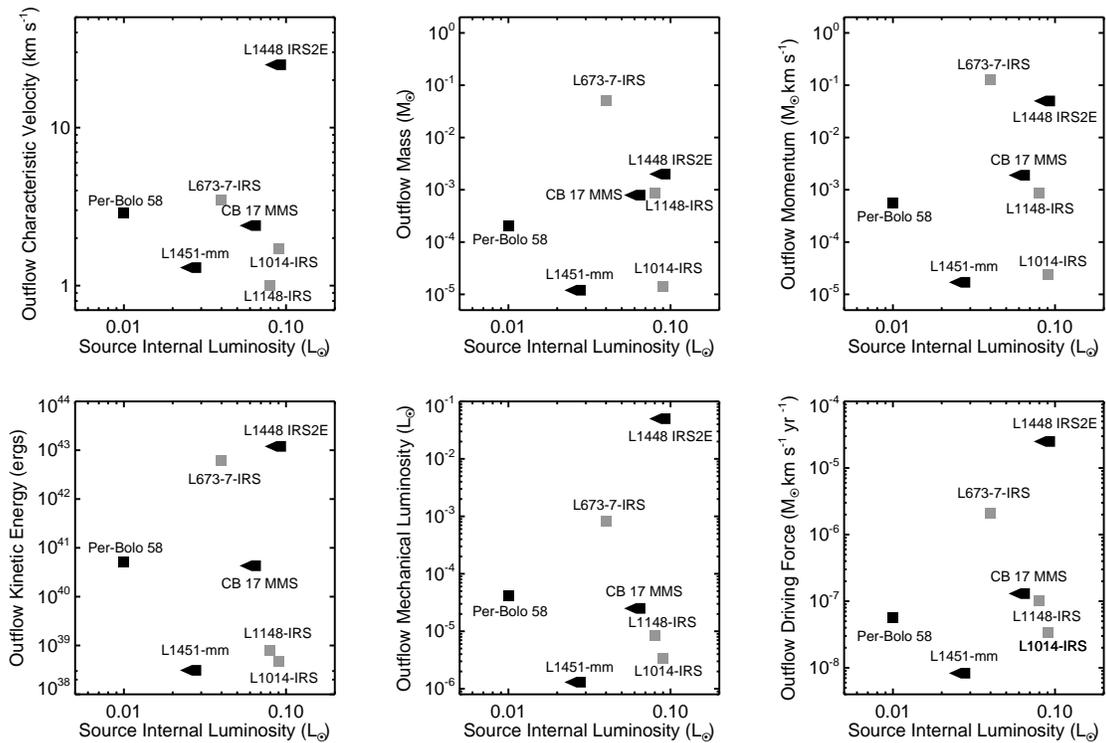}
\caption{\label{fig_outflow_props}Various outflow properties plotted vs. \lint\ for the sources listed in Table \ref{tab_outflow_others}.  Filled black symbols denote the candidate FHSCs and filled gray symbols denote the low luminosity protostars.  Each source is labeled, and the triangles pointing to the left denote upper limits in luminosity for three of the four candidate FHSCs (L1448 IRS2E, CB 17 MMS and L1451-mm).  Note that these upper limits are actually upper limits to \lbol\ rather than \lint\ and that \lint\ is likely less than \lbol, which will include a component from external heating.  However, without any far-infrared detections, it is not possible to separate internal and external components to \lbol\ for these sources.}
\end{figure*}

Table \ref{tab_outflow_others} lists the outflow properties of several other embedded sources with similarly low luminosities.  L1448 IRS2E is a dense core driving a molecular outflow with no associated \emph{Spitzer} infrared source and has been identified as a candidate FHSC by Chen et al.~(2010).  Similarly, the dense cores CB 17 MMS (X.~Chen et al.~2011, in preparation) and L1451-mm (Pineda et al.~2011) are associated with outflow detections (tentative in the case of CB 17 MMS), lack associated \emph{Spitzer} infrared sources, and have been identified as candidate FHSCs.  L1014-IRS is a very low luminosity ($\sim$ 0.09 \lsun; Young et al. 2004) Class 0 protostar driving a compact outflow detected only in interferometer observations (Bourke et al.~2005).  Similarly, L1148-IRS is a borderline Class 0/I protostar with \lint\ $\sim$ 0.10 \lsun\ also driving a compact outflow detected only in interferometer observations (Kauffmann et al.~2011).  Finally, L673-7-IRS is a very low luminosity ($\sim$ 0.04 \lsun) Class 0 protostar that is driving a large-scale outflow (Dunham et al.~2010).  Figure \ref{fig_outflow_props} plots these various outflow properties versus \lint\ for these six sources plus Per-Bolo 58.  Significant dispersion of at least one order of magnitude (and sometimes much more) is present in all properties, and no obvious correlation with luminosity is seen.  Per-Bolo 58, CB 17 MMS, and L1451-mm feature slow outflow velocities consistent with predictions by Machida, Inutsuka, \& Matsumoto (2008) for FHSC outflows, whereas L1448 IRS2E drives an outflow with a characteristic velocity (25 \kms) much faster than predicted for first cores.  However, the protostars L1014-IRS, L1148-IRS, and L673-7-IRS  also drive slow ($\leq 3.5$ \kms) outflows, and in general there is no clear separation between the properties of the outflows driven by the candidate FHSCs and by the very low luminosity protostars.

%\begin{landscape}
\begin{deluxetable*}{llllllll}
\tabletypesize{\tiny}
\tablewidth{0pt}
\tablecaption{\label{tab_outflow_others}Comparison of Outflow Properties}
\tablehead{\colhead{Quantity}             & \colhead{Per-Bolo 58} & \colhead{L1448 IRS2E\tablenotemark{a}} & \colhead{CB 17 MMS\tablenotemark{b}} & \colhead{L1451-mm\tablenotemark{c}} & \colhead{L1014-IRS\tablenotemark{d}} & \colhead{L1148-IRS\tablenotemark{e}} & \colhead{L673-7-IRS\tablenotemark{f}}}
\startdata
Source Internal Luminosity (\lint; \lsun)                & 0.012                & $<$0.1              & $<$0.07              & $<$0.03            & 0.09               & 0.10               & 0.04                \\
Lobe Size ($R_{\mathrm lobe}$; AU)                         & $6.2 \times 10^{3}$  & $9.6 \times 10^{3}$  & $8.0 \times 10^{3}$  & $5.5 \times 10^{2}$ & $5.4 \times 10^{2}$ & $1.8 \times 10^{3}$ & $4.5 \times 10^{4}$  \\
Outflow Mass ($M_{\mathrm flow}$; \msun)                   & $2.0 \times 10^{-4}$ & $2.0 \times 10^{-3}$ & $7.9 \times 10^{-4}$ & $1.2 \times 10^{-5}$ & $1.4 \times 10^{-5}$ & $8.5 \times 10^{-4}$ & $5.0 \times 10^{-2}$ \\
Outflow Momentum ($P_{\mathrm flow}$; \msun\ \kms)         & $5.6 \times 10^{-4}$ & $5.0 \times 10^{-2}$ & $1.9 \times 10^{-3}$ & $1.7 \times 10^{-5}$ & $2.4 \times 10^{-5}$ & $8.5 \times 10^{-4}$ & $1.3 \times 10^{-1}$ \\
Outflow Kinetic Energy ($E_{\mathrm flow}$; ergs)          & $5.1 \times 10^{40}$ & $1.2 \times 10^{43}$ & $4.3 \times 10^{40}$ & $3.1 \times 10^{38}$ & $4.8 \times 10^{38}$ & $8.0 \times 10^{38}$ & $6.1 \times 10^{42}$ \\
Outflow Luminosity ($L_{\mathrm flow}$; \lsun)             & $4.2 \times 10^{-5}$ & $5.0 \times 10^{-2}$ & $2.5 \times 10^{-5}$ & $1.3 \times 10^{-6}$ & $3.4 \times 10^{-6}$ & $8.5 \times 10^{-6}$ & $8.3 \times 10^{-4}$ \\
Outflow Force ($F_{\mathrm flow}$; \msun\ \kms\ yr$^{-1}$)  & $5.6 \times 10^{-8}$ & $2.5 \times 10^{-5}$ & $1.3 \times 10^{-7}$ & $8.3 \times 10^{-9}$ & $3.4 \times 10^{-8}$ & $1.0 \times 10^{-7}$ & $2.1 \times 10^{-6}$ \\
Characteristic Velocity ($v_{\mathrm flow}$; \kms)         & 2.9                  & 25                  & 2.4                 & 1.3                  & 1.7                  & 1.0                  & 3.5                 \\
Dynamical Time ($\tau_{\mathrm d}$; yr)                    & $1.0 \times 10^{4}$  & $1.8 \times 10^{3}$ & $1.4 \times 10^{4}$  & $2.0 \times 10^{3}$  & $7.0 \times 10^{2}$  & $8.5 \times 10^{3}$  & $6.1 \times 10^{4}$  \\
\enddata \\
\tablenotetext{a}{Chen et al.~(2010)}
\tablenotetext{b}{X.~Chen et al.~(2011, in preparation)}
\tablenotetext{c}{Pineda et al.~(2011)}
\tablenotetext{d}{Bourke et al.~(2005)}
\tablenotetext{e}{Kauffmann et al.~(2011)}
\tablenotetext{f}{Dunham et al.~(2010)}

\end{deluxetable*}
%\end{landscape}

While these findings argue against a distinct difference in evolutionary status between these two groups of objects, we caution that the sample size is extremely small, three of the four candidate FHSCs lack any mid and far-infrared detections and thus only have upper limits for their luminosities, and the outflow properties themselves are highly uncertain due to nonuniform assumptions about outflowing gas temperature, source inclination, optical depth, and methods of calculating properties.  This figure must be revisited as more sources (both candidate FHSCs and low luminosity protostars) with well-constrained inclinations are identified and their luminosities and outflow properties are calculated from uniform datasets with consistent methods, and as additional simulations fully explore predicted differences between the kinematics, energetics, and morphologies of outflows driven by first cores and protostars.

A significant unknown in our analysis of the kinematics and morphology of the Per-Bolo 58 outflow and comparison to the predictions of Machida, Inutsuka, \& Matsumoto (2008) is the source inclination.  As discussed above, we are unable to set strong constraints on the inclination from the outflow morphology.  Both relatively edge-on and relatively face-on configurations are possible, with only extreme edge-on ($\ga$ 80\degree) and face-on ($\la$ 10\degree) inclinations ruled out\footnote{Enoch et al.~(2010) also ruled out inclinations $\ga$ 80\degree\ based on their radiative transfer models.}.  If Per-Bolo 58 is viewed along a relatively edge-on line-of-sight, the true velocity of the outflow would increase beyond that expected for FHSCs.  For example, an inclination of 80\degree\ would increase the outflow velocity to 16.5 \kms.  On the other hand, if Per-Bolo 58 is viewed along a relatively face-on configuration, the true outflow velocity would be similar to the characteristic radial velocity of 2.9 \kms\ and the observed SED would only be consistent with FHSC models and not protostellar models (Enoch et al.~2010).  However, the extent-to-width ratio would increase substantially beyond that predicted by Machida, Inutsuka, \& Matsumoto (2008) for FHSC outflows and the dynamical time would increase beyond the maximum expected lifetime of first cores.  For example, at an inclination of 10\degree\ the outflow velocity would remain at 2.9 \kms\ but the lobe size would increase to $3.5 \times 10^4$ AU, the extent-to-width ratio would increase to $\sim$ 20, and the dynamical time would increase to $5.7 \times 10^4$ yr.

Clearly, knowledge of the source inclination would aid in determining the evolutionary status of Per-Bolo 58.  Inclination constraints could be provided by very deep near- or mid-infrared images that detect scattered light cones from the outflow cavities, combined with radiative transfer model images at different inclinations (e.g., Huard et al.~2006; Terebey et al.~2006).  They could also be obtained through very high-resolution interferometer observations separated by several years that measure the proper motion of the outflow (e.g., Lee et al.~2009). 

If Per-Bolo 58 turns out to be a protostar rather than a FHSC, our current understanding of the protostellar luminosity distribution would be incomplete.  Evans et al.~(2009) and Enoch et al.~(2009) identified 112 embedded protostars in the \emph{Spitzer} c2d Legacy dataset and showed that their luminosities vary between $\sim 0.05 - 50$ \lsun, with the distribution skewed towards low luminosities (59\% have $L < 1.6$ \lsun).  In a closely related investigation, Dunham et al.~(2008) presented a detailed study of the low end of the protostellar luminosity distribution and found that the distribution, which generally rises to lower luminosities, flattens out below 0.1 \lsun\ and appears to remain flat down to their sensitivity limit of 0.004 \lsun\ at 140 pc (or 0.013 \lsun\ at 250 pc, the distance to Perseus).  They noted that the fact that sources were found all the way down to their sensitivity limit implied there could be a population of objects at even lower luminosities.  Per-Bolo 58 as a protostar would be a member of such a population.  Future work must address this issue by further study of the low end of the luminosity distribution of embedded sources through a combination of sensitive interferometer outflow surveys and very deep \emph{Herschel} and \emph{James Webb Space Telescope} infrared surveys directed towards cores currently classified as starless to identify other objects similar to Per-Bolo 58, L1448 IRS2E, CB 17 MMS, and L1451-mm.

\section{Summary}\label{sec_summary}

In this paper we have presented new 230 GHz Submillimeter Array observations of the candidate first hydrostatic core Per-Bolo 58 and reported the detection of a continuum source and a bipolar molecular outflow, both centered on the position of the infrared source embedded within this core.

The continuum detection has a total flux density of $26.6 \pm 4.0$ mJy, and from this detection we calculate a mass of $0.11 \pm 0.05$ \msun\ and a mean number density of $2.0 \pm 1.6 \times 10^7$ cm$^{-3}$.  The total mass of the Perseus Bolo 58 core is approximately 1 \msun, indicating that the SMA filters out all but the compact, dense inner region.  Fits to the visibility amplitudes suggest the presence of an unresolved component with a total flux density of 4.7 mJy, but observations in more extended configurations providing longer baselines are needed to confirm the presence of this unresolved component and determine whether it is tracing the dense inner envelope or a circumstellar disk.

The outflow is detected in the \cojtwo\ observations and extends along a nearly due east-west axis at a position angle of 85\degree\ (measured east from north).  The blue lobe extends to the east and the red lobe extends to the west with very little overlap between the lobes.  The outflow is slow (characteristic velocity of 2.9 \kms), shows a jet-like morphology (opening semi-angles $\sim$8\degree\ for both lobes), and extends to the edges of the primary beam.  The morphology and kinematics of this outflow are partially consistent with predictions for outflows driven by both first hydrostatic cores and protostars, and do not fully distinguish between the two possibilities.  We calculate the kinematic and dynamic properties of the outflow in the standard manner and compare them to several other protostars and candidate first hydrostatic cores with similarly low luminosities, and we show there are no clear trends between these properties and source luminosity, and no clear separation in properties between outflows driven by candidate first hydrostatic cores and very low luminosity protostars.

While Per-Bolo 58 is a very strong candidate to be a first core, ultimately we are still unable to determine the true evolutionary status of this source; it could be either a first hydrostatic core or a very low luminosity protostar.  We have discussed the evidence in support of both possibilities, as well as the implications for our current understanding of protostellar luminosities if it is in fact a protostar.  We have outlined future observations needed to better determine the physical and evolutionary properties of this source, including longer baseline continuum data and source inclination constraints provided by either deep infrared images or very high-resolution interferometer observations, and we have also discussed the need for further study of the low end of the protostellar luminosity distribution through both sensitive interferometer outflow surveys and very deep \emph{Herschel} and \emph{James Webb Space Telescope} infrared surveys directed towards cores currently classified as starless.

\acknowledgements
The authors extend their gratitude to the anonymous referee for comments and suggestions that have improved the quality of this work.  We also thank the SMA staff for executing these observations as part of the queue schedule, and Charlie Qi for his assistance with MIR and MIRIAD.  This work is based primarily on observations obtained with The Submillimeter Array, a joint project between the Smithsonian Astrophysical Observatory and the Academia Sinica Institute of Astronomy and Astrophysics and funded by the Smithsonian Institution and the Academia Sinica.  This research has made use of NASA's Astrophysics Data System (ADS) Abstract Service and of the SIMBAD database, operated at CDS, Strasbourg, France.  Support for this work was provided by the NSF through grant AST-0845619 to H.G.A.  T.L.B.~acknowledges support from NASA Origins grant NXX09AB89G.

%%%%%%%%%%%%%%%%%%%%%%%%%%%%%%%%%%%%%%%

%\clearpage

%%%%%%%%%%%%%%%%%%%%%%%%%%%%%%%%%%%%%%%

%\clearpage

%%%%%%%%%%%%%%%%%%%%%%%%%%%%%%%%%%%%%%%

%\clearpage

\end{document}